\documentclass[namedreferences]{SolarPhysics}
\usepackage[optionalrh]{spr-sola-addons} % For Solar Physics 
\usepackage{graphicx}                    % For eps figures, newer & more powerfull
\usepackage{color}                       % For color text: \color command
\usepackage{url}                         % For breaking URLs easily trough lines
\usepackage{solaheader}                  % Use this for preprint servers only

\newcommand{\solareceiveddate}{10 Jan 2010} % Remove the leading % and fill in the values when known; solaheader must be used
\newcommand{\solaaccepteddate}{14 June 2010}
\newcommand{\solaonlinedate}{13 July 2010}
\newcommand{\doi}{10.1007/s11207-010-9591-7}
\newcommand{\etal}{{\it et al.}}
\newcommand{\rng}{{\,--\,}}
\newcommand{\degree}{$^{\circ}$}

\newcommand{\logten}{log$_{10}$}
\newcommand{\logt}{log$_{10} T$}
\newcommand{\arcsec}{$^{\prime\prime}$} %% arcsecond symbol: 0''
\newcommand{\feone}   {{\rm Fe {\sc ix} 171.073 \AA}}
\newcommand{\fefour}  {{\rm Fe {\sc x} 174.531 \AA}}
\newcommand{\feeighty}{{\rm Fe {\sc xi} 180.408 \AA}}
\newcommand{\feeight} {{\rm Fe {\sc xi} 188.232 \AA}}
\newcommand{\fethree} {{\rm Fe {\sc xii} 193.509 \AA}}
\newcommand{\fefive}  {{\rm Fe {\sc xii} 195.119 \AA}}
\newcommand{\fetwootwo}   {{\rm Fe {\sc xiii} 202.044 \AA}}
\newcommand{\fetwoothree}{{\rm Fe {\sc xiii} 203.828 \AA}}
\newcommand{\feeleven}{{\rm Fe {\sc xiv} 211.318 \AA}}
\newcommand{\fefiftytwo}{{\rm Fe {\sc xiv} 251.956 \AA}}
\newcommand{\fesixtyfive}{{\rm Fe {\sc xiv} 264.790 \AA}}

\begin{document}

\begin{article}

\begin{opening}

\title{EUV Spectra of the Full Solar Disk: Analysis and Results
of the {\it Cosmic Hot Interstellar Plasma Spectrometer} (CHIPS)}

%%%%%%%%%%%%%%%%%%%%%%%%%%%%%%%%%%%%%%%%%%%%%%%%%%%
%% Authors Names
%
\author{M.~M.~\surname{Sirk}$^{1}$\sep
        M.~\surname{Hurwitz}$^{1}$\sep
        W.~\surname{Marchant}$^{1}$      
       }

%%%%%%%%%%%%%%%%%%%%%%%%%%%%%%%%%%%%%%%%%%%%%%%%%%%
%% Runningheads
%
\runningauthor{M.~M.~Sirk \etal}
\runningtitle{CHIPS Solar EUV Spectra}

%%%%%%%%%%%%%%%%%%%%%%%%%%%%%%%%%%%%%%%%%%%%%%%%%%%
%% Affilations 
%
  \institute{$^{1}$ Space Sciences Laboratory, University of California, Berkeley, CA, 94720 USA
                     email: \url{sirk@ssl.berkeley.edu} \url{markh@ssl.berkeley.edu} \url{marchant@ssl.berkeley.edu} \\
} 

%%%%%%%%%%%%%%%%%%%%%%%%%%%%%%%%%%%%%%%%%%%%%%%%%%%
%%% Abstract 
\begin{abstract}

We analyze EUV spectra of the full solar disk from the
{\it Cosmic Hot Interstellar Plasma Spectrometer} (CHIPS) spanning
a period of two years.
The observations were obtained via a fortuitous off-axis light path in
the 140\rng275 \AA\ passband.
The general appearance of the spectra remained relatively stable over
the two-year time period,
but did show significant variations of up to 25\% between two sets of
Fe lines that show peak emission at 1 MK and 2 MK.
The variations occur at a measured period of 27.2 days and are
caused by regions of hotter and cooler plasma rotating into,
and out of, the field of view.
The CHIANTI spectral code is employed to determine plasma temperatures,
densities, and emission measures.
A set of five isothermal plasmas fit the full disk spectra well.
A 1\rng2 MK plasma of Fe contributes 85\% of the total emission in
the CHIPS passband.
The standard Differential Emission Measures (DEMs)
supplied with the CHIANTI package do not fit the CHIPS
spectra well as they over-predict emission at temperatures
below \logt = 6.0 and above \logt = 6.3.
The results are important for cross-calibrating TIMED, SORCE,
SOHO/EIT, and CDS/GIS,
as well as the recently launched {\it Solar Dynamics Observatory}.

\end{abstract}

%%%%%%%%%%%%%%%%%%%%%%%%%%%%%%%%%%%%%%%%%%%%%%%%%%%
%% Keywords
%

\end{opening}
%-------------------------------------------------

%%%%%%%%%%%%%%%%%%%%%%%%%%%%%%%%%%%%%%%%%%%%%%%%%%%
%% Sections
%

\section{Introduction}
     \label{S-Introduction} 

%Knowledge of solar variability and total irradiance is essential for
%understanding terrestrial atmospheric processes.
%EUV radiation photoionizes neutral elements creating the ionosphere.
%Secondary processes then heat the thermosphere.
%(More Needed here????)

The solar EUV irradiance is the primary source of energy and ionization of the
Earth's upper atmosphere.  Variations in the EUV flux are responsible for
dramatic changes in the density of neutrals and ions, affecting satellite
drag forces in near-Earth orbit by up to a factor of ten and modifying the
propagation of radio frequency signals \cite{Fuller-Rowell04}.

Solar high spectral-resolution EUV spectra were first obtained via sub-orbital
rocket flights \cite{Behring72,Malinovsky73,Thomas94},
and {\it SkyLab} \cite{Dere78}.
Collectively, these early missions discovered over 400 emission lines
from 17 elements between 50 \AA\ and 600 \AA,
indicating the presence of a 1 to 2 MK plasma dominated by emission from Fe.

%The Solar EUV Rocket Telescope and Spectrograph (SERTS) found 269
%emission lines (mostly of Fe) between 171\AA\ and 450\AA\
%from a slit spectrogram of an active region \cite{Thomas94}.

The {\it Solar and Heliospheric Observatory/EUV Imaging Telescope}\break
(SOHO/EIT)
obtains full-disk images in four $\approx$10 \AA\ (FWHM) passbands
centered at 171, 195, 285, and 304 \AA.
These pictures show where EUV emission originates, however,
as concluded by \inlinecite{Hock08},
it is essentially impossible to accurately convert the broad-band photometry to
physical units without simultaneous knowledge of the solar spectrum.

SOHO/CDS/GIS has one channel (150 to 220 \AA) that overlaps CHIPS, but
shows spectral ``ghosts'' and temporal variations in throughput that require
careful calibration \cite{Kuin07} to provide acurate line intensities
and ratios. %%% more ....

The {\it Hinode/Extreme Ultraviolet Spectrograph} (EIS) is a high spectral-
and spatial-resolution instrument that provides very detailed
temperature and density maps,
but can only see a small strip (1 or 2\arcsec\ wide by 512\arcsec\ long)
of the Sun \cite{Young09}.

Each of these instruments has its advantages and limitations.
At one extreme are the high spectral- and spatial-resolution spectrographs
that are not well suited for total solar irradiance;
at the other are the broad-band photometers which, even if
accurately calibrated, cannot provide details of the solar spectrum.
Rocket flights typically provide only a four-minute snapshot, and thus cannot
address temporal variations.

Great progress in determining the EUV contribution to total solar irradiance
has been made by the {\it Thermosphere, Ionosphere, Mesosphere, Energetics,
and Dynamics} (TIMED), and {\it Solar Radiation and Climate Experiment}
(SORCE) missions \cite{Woods08}.
The daily data provided by these instruments are well calibrated,
broad-band (width 70 to 100 \AA) photodiode
measurements, which are used to scale reference CHIANTI \cite{Landi06,Dere97}
model spectra (quiet Sun, active region, coronal hole, and flare)
that depend upon assumed
Differential Emission Measures (DEMs) and out-of-band proxies.
We refer to the SORCE XPS Level 4 Version 10 model (outlined in detail
in Woods \etal, 2008) as XPSL4 hereafter.

Here we present full-disk solar spectra from the
{\it Cosmic Hot Interstellar Plasma Spectrometer} (CHIPS)
covering a period of two years.
The data from the SORCE, TIMED/SEE, and SOHO/EIT missions complement those
of CHIPS.
The first two provide the total irradiance,
SOHO/EIT shows where the EUV emission originates,
and CHIPS provides spectra from which temperatures and densities
may be determined as well as
provide the spectral shape between 140 and 270 \AA\ required
to improve the calibration of the contemporaneous SORCE/XPS and TIMED/SEE
photodiode measurements, and the SOHO/EIT broadband full-disk images.
A recent flight of the rocket prototype
{\it Solar Dynamics Observatory} (SDO)/{\it Extreme Ultraviolet Variability
Experiment} (EVE) \cite{Woods09} provided
an EUV spectrum with similar passband and resolution as CHIPS.
This spectrum is utilized to cross-check the quality of
the solar observations and the calibration of the CHIPS throughput.
An uncalibrated, mean CHIPS spectrum was first presented by
\inlinecite{Hurwitz06},
and compared to laboratory plasmas by \inlinecite{Lepson08}.

In this work we explore to what extent the solar corona can be described
by a set of simple plasmas, and how these may be used in modeling the
spectral shape of the solar irradiance. 
We present both raw and flux calibrated spectra,
quantify the spectrometer's response to off-axis solar illumination,
perform CHIANTI plasma modeling,
and outline the contents of a publicly available archive.

\section{CHIPS Instrument}
     \label{S-Instrument}

The CHIPS satellite is a 
NASA University Explorer mission
devoted to diffuse background spectroscopy of the interstellar medium
at moderate resolution ($\lambda/ (\Delta \lambda$FWHM)$\approx 120$)
in the EUV passband 90\rng260 \AA.
NASA launched CHIPS from Vandenberg AFB on 12 January 2003 as a secondary
payload on a Delta II rocket into a circular polar orbit with an inclination
of 94\degree\ and an altitude of 590 km.

The three-axis stabilized satellite consists of a 35-kg spacecraft built by\break
SpaceDev, Inc., and a 25-kg science instrument provided by the University of
California, Berkeley.
%%%  The integrated assembly is about 100 by 100 by 30 cm.
The spectrograph consists of a micro-channel plate detector placed at
the focus of six varied-line-spacing cylindrically curved
diffraction gratings.
Mounted 8.8 mm in front of the detector face is a filter frame which
holds the vacuum deposited Al, Zr, and polyamide-boron thin film 
($\approx$ 1000 \AA\ thick) filters, which suppress higher diffracted orders as
well as longer-wavelength scattered and stray light.
Unlike traditional imaging spectrographs, the CHIPS design does not employ a
telescope to gather light.
Regardless of a source's position within the field of view, light
of a particular wavelength is directed at a particular dispersion
angle to the micro-channel plate detector.
A single spectrum of the entire 4\degree\ $\times$ 26\degree\ field of view
may thus be obtained simultaneously. 
The opto-mechanical design of the instrument is outlined in
\inlinecite{Sholl03a},
the detector characteristics in \inlinecite{Marckwordt03},
the calibration and in-flight performance in \inlinecite{Sirk03},
and the primary mission science results in \inlinecite{Hurwitz05}.

%% Figure grating_fan.eps

 \begin{figure} 
\centerline{\includegraphics[width=0.75\textwidth,clip=]{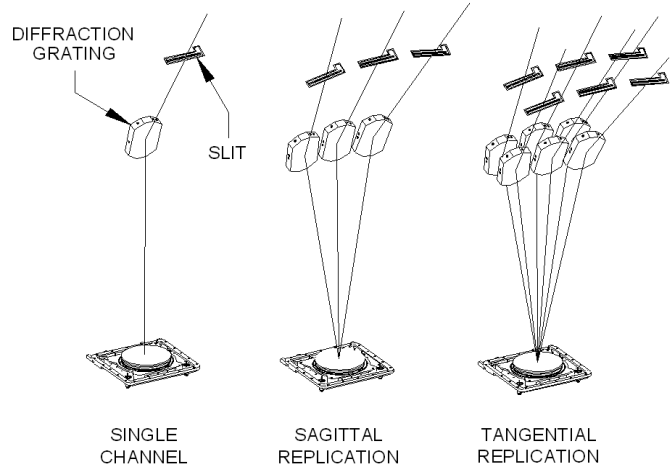}}
 \caption{Schematic showing relative locations of the six slit plates
 and diffraction gratings employed by CHIPS. Left: central channel. Middle: 
 non-mirrored channels. Right: non-mirrored and mirrored channels (rear row).}
 \label{F-grating}
 \end{figure}

\begin{figure} 
\centerline{\includegraphics[width=1.0\textwidth,clip=]{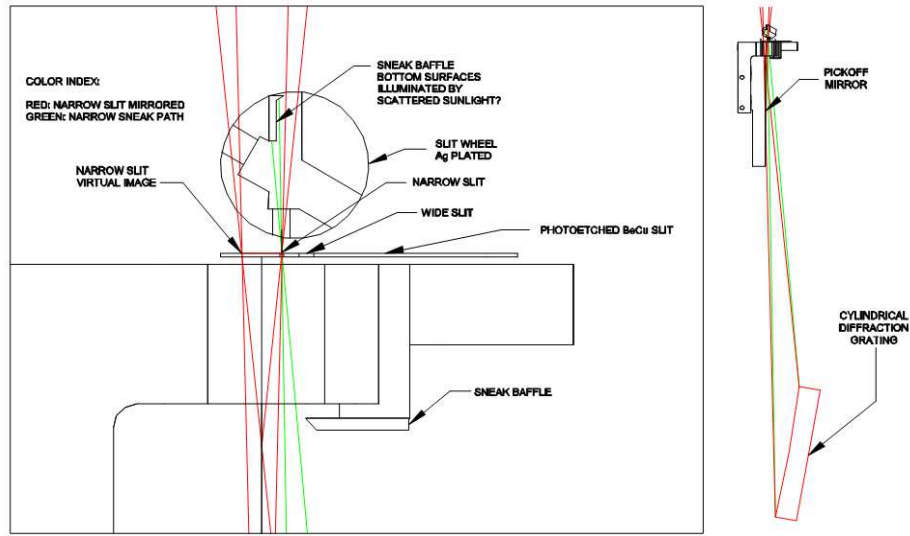}}
 \caption{Edge-on views of slit and slit-wheel shutter mechanism showing
 direct and scattered light paths. Left:
 field of view of direct path and virtual image of direct path (red),
 and field of view of scattered path off baffle (green). Right: general
 view showing location of slit wheel, pickoff mirror, and diffraction grating.
 Notice that the scattered path does not intersect the pickoff mirror.}
 \label{F-slit_wheel}
 \end{figure}

Each diffraction grating is equipped with a pair of adjacent, fixed, entrance
slits, 0.25 and 1.0 mm wide (Figure~\ref{F-grating}).
In front of each slit pair is a rotary
mechanism that allows diffuse light to illuminate the grating through
one slit or the other (or can be set to block both).  Between the slits and
three of the gratings are flat grazing-incidence mirrors that serve to
co-align the field of view of those spectral channels with their
non-mirrored counterparts (see Figure~\ref{F-slit_wheel}, right).
During normal science observations,
the instrument boresight is constrained to be
far from the Sun, so that the rotary 
mechanism and entrance slits are shaded from direct sunlight.
When several comets presented themselves as
targets of opportunity \cite{Sasseen06}, we relaxed this constraint 
and allowed the instrument boresight to
encroach within tens of degrees of the Sun.
The resulting spectra showed features that seemed not
to originate from the comets, were far too bright to arise in
the interstellar medium, and did not (without shifting) match
the reflected solar spectrum that had been recorded during
observations of the full Moon.

We ultimately discovered that in certain geometries, scattered sunlight 
could pass through an entrance slit, avoid the flat mirror,
and strike a diffraction grating at an angle different from the nominal
14\degree\ incidence.  Ray-trace analysis by us and \inlinecite{Sholl05}
%%(CHIPS Technical Memorandum 24 October 2005)
confirmed that light scattered from the back side of a baffle could follow
this path to the detector, producing a spectrum shifted by $\approx$ 11 \AA\ 
from the nominal wavelength solution (see Figure~\ref{F-slit_wheel}, left).
After much experimentation with the Sun angle and slit orientation
(angle of slit axis with respect to the ecliptic plane), a configuration
was found that nominally showed a single channel solar spectrum.
We refer to this path as the ``light-leak.''
The light-leak has no adverse effect on ordinary science observations.
Its throughput is six orders of magnitude lower than
for emission within the direct field of view, and diffuse emission
comprising the main science target is so faint as to be barely detectable.
The light-leak %%, a term with pejorative connotations,
in this case provides a serendipitous way of
measuring the EUV spectrum of light integrated from the solar disk.

%% Figure slit_wheel.eps

\section{Observations}
     \label{S-Observations}

The CHIPS satellite performed 1450 observations of the Sun from
03 April 2006 to 05 April 2008,
a period of generally declining activity following the solar maximum of 2001.
The satellite was decommissioned on 14 April 2008.
Typical observations ranged from 1 to 15
minutes which yielded spectra of  roughly $10^5$ total counts each.
Occasionally, up to 15 such observations per day were performed.
However for a variety of causes, including passages through the
South Atlantic Magnetic Anomaly, passages through the electron belts at
high magnetic latitudes,
targets of opportunity, precession of the orbit into full sunlight,
and satellite shutdowns, the cadence of observations
is far from uniform.

\begin{figure} 
 \centerline{\includegraphics[width=0.75\textwidth,clip=]{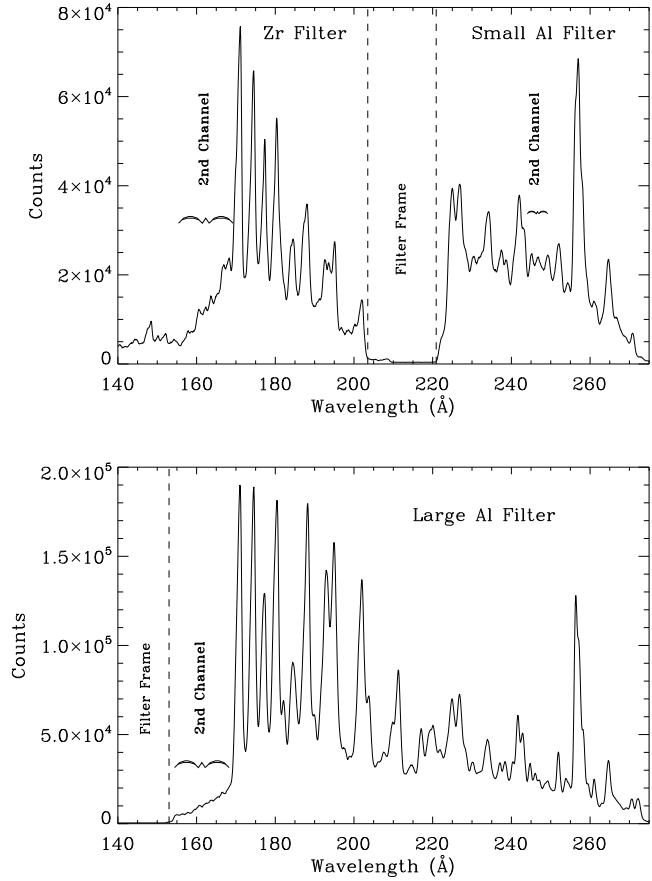}}
 \caption{Raw CHIPS spectra binned at 0.11 \AA\ for the Zr/small Al filter
 (top) and the large Al filter (bottom).  Wavelength regions of filter-frame
 occlusion and second channel contamination
 (described in Section~\ref{S-multiple-channels}) features are indicated.}
 \label{F-rawspec}
 \end{figure}

We present raw spectra integrated over the two-year period for
the upper Zr/Al filters, and the lower large Al filter in
Figure~\ref{F-rawspec}.
The brightest features are from Fe {\sc ix} through Fe {\sc xv}.
Also evident is a continuum of in-band scattered light
from the diffraction gratings superimposed with second (and possibly
third) channel features 
(discussed in Section~\ref{S-multiple-channels}),
which occur more frequently, and to a greater degree, in
the upper Zr/Al filter.
Particle background (primarily high energy electrons) is
negligible as evinced by the near zero count rates in
the shadows of the filter frames. 

The general appearance of the daily spectra remained relatively stable.
The emission lines labeled in Figures~\ref{F-alspec} and ~\ref{F-zrspec}
did not appear and disappear, but were visible at all times, even during
the quiet Sun period of April 2008.
During 2006, CHIPS observed an M-class flare
(27 April 2006 15:22 UT M7.9), and just missed a second one by 24 hours
(06 July 2006 08:13 UT M2.5).
Spectra extracted in the hours just prior to and during the flares showed
no discernible differences in line ratios, indicating that moderate intensity
flares do not necessarily produce large changes
in the shape of the EUV spectrum.
We present a more detailed analysis in Section~\ref{S-flare}.

%% Figure raw_spec.eps 

\section{Calibrations}
  \label{S-Calibrations} 
     \subsection{Calibration Strategy} %% and Preprocessing}
       \label{S-calibration-strategy}

Each observation is processed by our standard science pipeline
and spectra extracted as described by \inlinecite{Hurwitz05}.
Because the solar data are obtained via the fortuitous light-leak,
the ground-based calibrations for wavelength scale and throughput are
not directly applicable.
To analyze the spectra quantitatively, the instrument required
an-{\it ad hoc} recalibration using the solar data itself.
The strategy is one of iteration: as various instrumental effects become
known, the entire data set is reprocessed to produce the highest-quality
spectra.

     \subsection{Reference Spectra}
       \label{S-reference-spectra}

The first step is to create a provisional reference spectrum 
for each filter set by simply adding up all the observations.
Next, each individual spectrum is shifted in the dispersion direction
until it best matches (by cross correlation) the provisional
reference spectrum, thus placing all data onto a common dispersion grid.
The shifts proved to be minor, typically 0 to 0.2 \AA, demonstrating
that the spectrograph is essentially free from flexure problems.
Finally, all of the shifted spectra are co-added a second time to create
improved reference spectra for both the Al filter and the Zr/Al filter panels.

     \subsection{Multiple Channel Contamination}
       \label{S-multiple-channels}

In spite of our efforts to best configure the satellite orientation
to produce a single-channel spectrum,
multiple channel contamination is still evident in the majority of
pointings and is quite variable in intensity from one observation
to the next.
The contamination is not the result of second-order grating features,
but rather, is caused by scattered light from a second (or third)
slit and grating pair striking the detector (see Figure~\ref{F-grating}).
The contamination is manifest as additional features
shortward of \feone\ in Figure~\ref{F-rawspec} in both the Zr and Al filters.
The likely cause for the observed variations is that
small errors in the pointing knowledge of the satellite 
of $\approx 1.0$\degree\ lead to
large changes in the illumination of the slits, with each channel responding
differently with regard to wavelength scale and throughput.
Efforts to decouple the channels proved futile.
We found, for example, that a multi-channel spectrum is not simply the
linear combination of scaled and shifted reference spectra, even if
the reference spectrum is constructed from contamination-free spectra.

Fortunately, the degree of second channel contamination is in general small and
is easily quantified by comparing count rates on either side of the absorption
edge of the Al filter at 170 \AA.
The effective area of the Al filter determined by ground calibrations
\cite{Sirk03} show that the sensitivity just shortward of 170 \AA\ is
about 13\% of the 170\rng180 \AA\ region.
Since second-channel contamination
is shifted bluewards relative to the ``good'' channel, any measured flux
shortward of 170 \AA\ that is greater than expected is a direct indication
of contamination.
We calculate a contamination fraction for each observation as the ratio of the
average counts per \AA\ between 155 \AA\ and 167 \AA\ divided by the 
average counts per \AA\ of the four brightest iron lines
\feone, \fefour, \feeighty, and \feeight\ for the Al filter.
The contamination ranges from 0 to 41\% with a median value of 6\%.
By selecting appropriate contamination thresholds, the unwanted light
may be greatly reduced, or even eliminated, albeit at the expense
of rejecting a portion of the observations.
For the lightcurve, line ratio, and line-intensity correlation analysis
described in Section~\ref{S-lightcurves},
we choose a threshold of 12\%,
which results in a rejection of about 40\% of the observations.
This is a compromise between accepting a small amount of unwanted
light and rejecting large quantities of the data.
For the CHIANTI plasma modeling (Section~\ref{S-modeling}), 
we apply a more stringent threshold of
8\% when fitting the Zr filter data since they show a greater degree
of contamination, and because there are Ni 
emission features of critical interest shortward of 170 \AA .

     \subsection{Background Subtraction}
       \label{S-background}

\begin{figure} 
 \centerline{\includegraphics[width=1.1\textwidth,clip=]{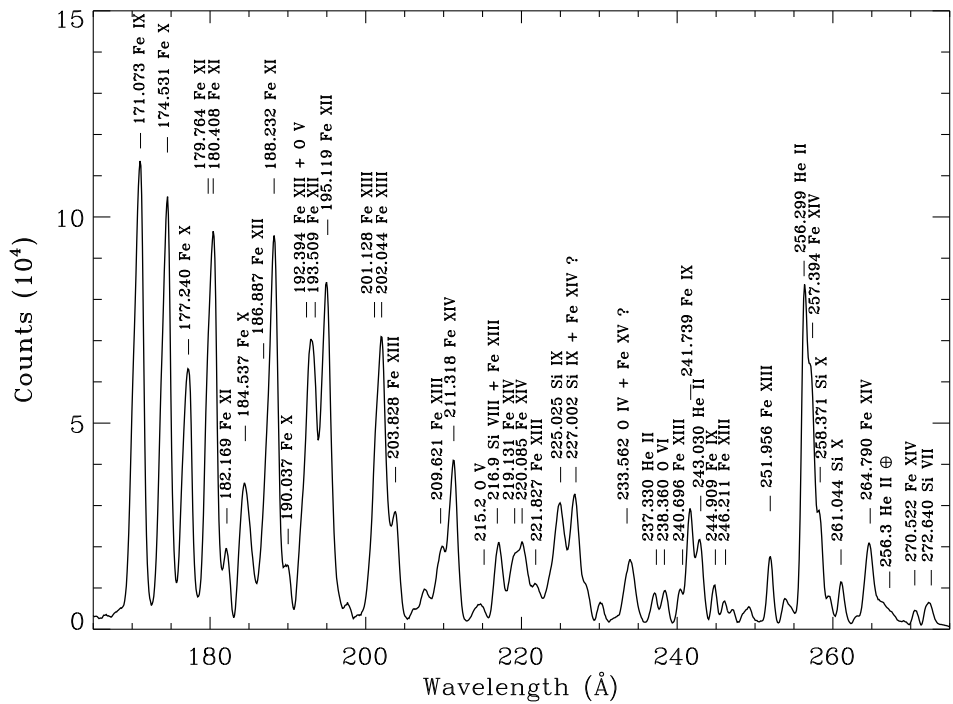}}
%\caption{\cite{Landi06, Dere97}}
 \caption{Cleaned, time-averaged, background-subtracted
 Al-filter spectrum binned at 0.11 \AA.
 Line identifications and wavelengths are from the CHIANTI database
(Landi \etal 2006; Dere \etal 1997).}
  \label{F-alspec}
 \end{figure}

\begin{figure} 
 \centerline{\includegraphics[width=1.1\textwidth,clip=]{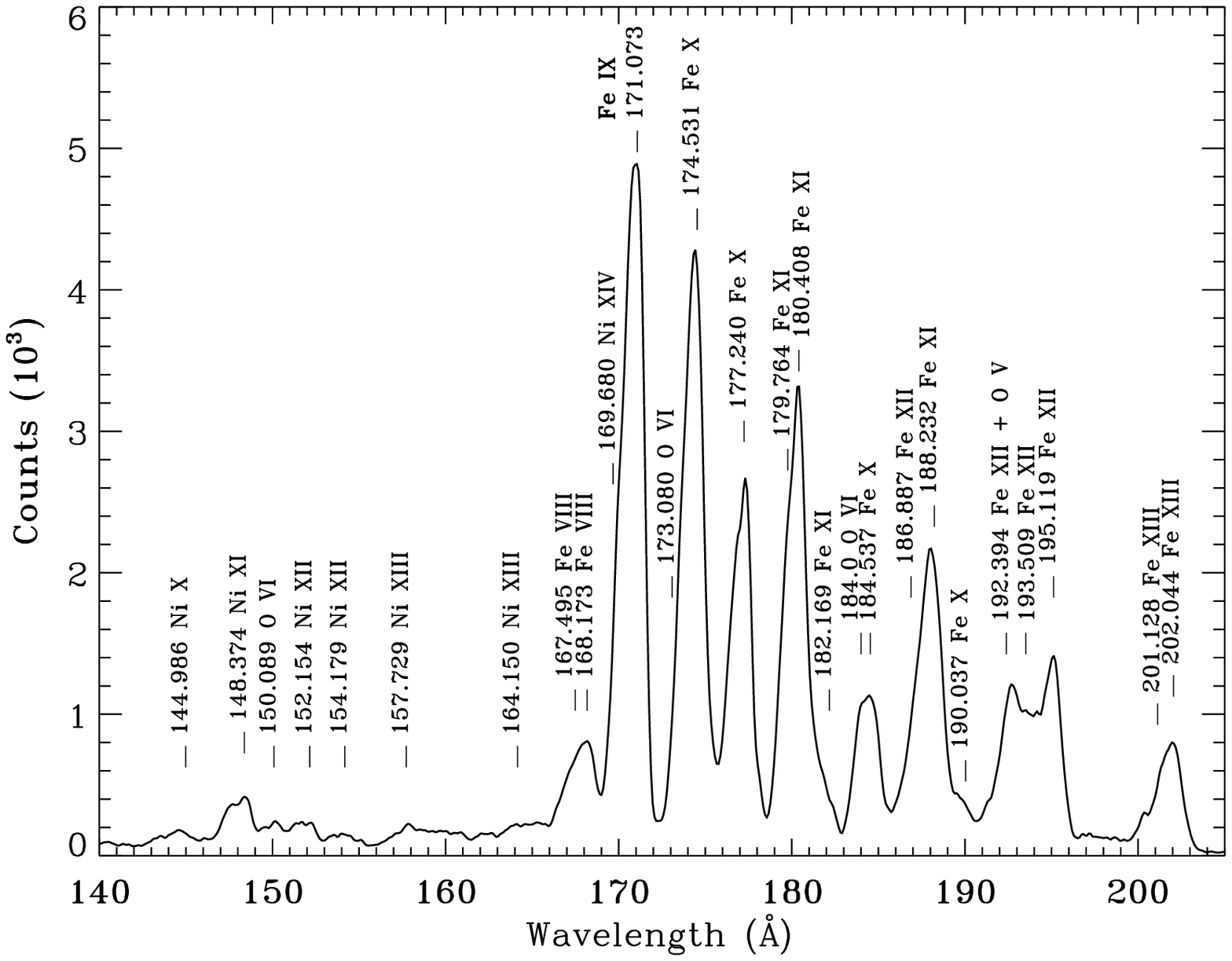}}
 \caption{Cleaned, time-averaged, background-subtracted
 Zr-filter spectrum binned at 0.11 \AA.
 Line identifications
 for Ni {\sc x} and Ni {\sc xi} are from Mewe \etal, 1985.}
 \label{F-zrspec}
 \end{figure}

The CHIPS solar-spectra background is dominated by in-band scattering from the 
diffraction gratings.
High-energy particles, second-order features,
and out-of-band scattering from geocoronal emission lines are all negligible.
Initial mean backgrounds are determined from the reference spectra by
simply drawing by hand. 
Provisional CHIANTI model fits
(specifically outlined in Section~\ref{S-modeling})
are performed and the backgrounds adjusted
until reasonable fits are found. 
In Figures~\ref{F-alspec} and ~\ref{F-zrspec} we present the
time-averaged
background-subtracted spectra with negligible second-channel contamination for
the Al and Zr filters, respectively.
These filtered spectra along with their respective backgrounds 
are saved as new reference spectra, and the steps
outlined in Sections~\ref{S-reference-spectra} and~\ref{S-multiple-channels}
repeated once more to create a database with all spectra on a common
grid.  For analysis, spectral subsets may be easily created by selections
based on date, count rate, contamination fraction,
line ratios, exposure time, {\it etc.} 

%% Figure alspec.eps 

%% Figure zrspec.eps 

     \subsection{Wavelength Scale}
       \label{S-wavelength-scale}

The dispersion of the CHIPS spectrograph is known to be nearly linear
\cite{Sirk03} with an average value of 0.11 \AA\ per pixel.
Ray-trace analysis also shows that small translations of the slit
position cause nearly constant shifts in wavelength of spectral features.
Because the solar spectra light-leak is along a unforeseen path, the
observed spectral shifts are effectively caused by a slit translation.
Thus, the solar wavelength dispersion is also nearly linear, and
is shifted about 11 \AA\ bluewards relative to the ground-based
wavelength solution. 
Twenty lines are identified in the Al and Zr/Al cleaned reference spectra
using the CHIANTI V5.2.1 database \cite{Landi06,Dere97}
and a quadratic fit performed to determine wavelength
as a function of detector {\it x} position.
Each fit shows an RMS residual of 0.08 \AA.
The resolution of the reference spectral features is about 1.2 \AA\ FWHM,
or $R=140$ at 170 \AA.

\section{Analysis}
  \label{S-Analysis} 

The CHIPS EUV spectra are for the full solar disk, and thus a melange %mixture 
of emission from active regions, quiet Sun, small flares, and coronal holes.
Both temperature and density in these plasmas range over two orders
of magnitude.
In Section~\ref{S-lightcurves},
we investigate the observed temporal variations from one
observation to the next and compare the CHIPS data to the sunspot number,
and the SORCE XPS L4 V10 spectral model flux
(described in Section~\ref{S-Introduction}).
In Section~\ref{S-modeling} we 
determine values for temperature and density that best describe
the full-disk for time-averaged spectra.

    \subsection{Lightcurves, Line Ratios, and Correlations}
      \label{S-lightcurves} 

\begin{figure} 
 \centerline{\includegraphics[width=1.1\textwidth,clip=]{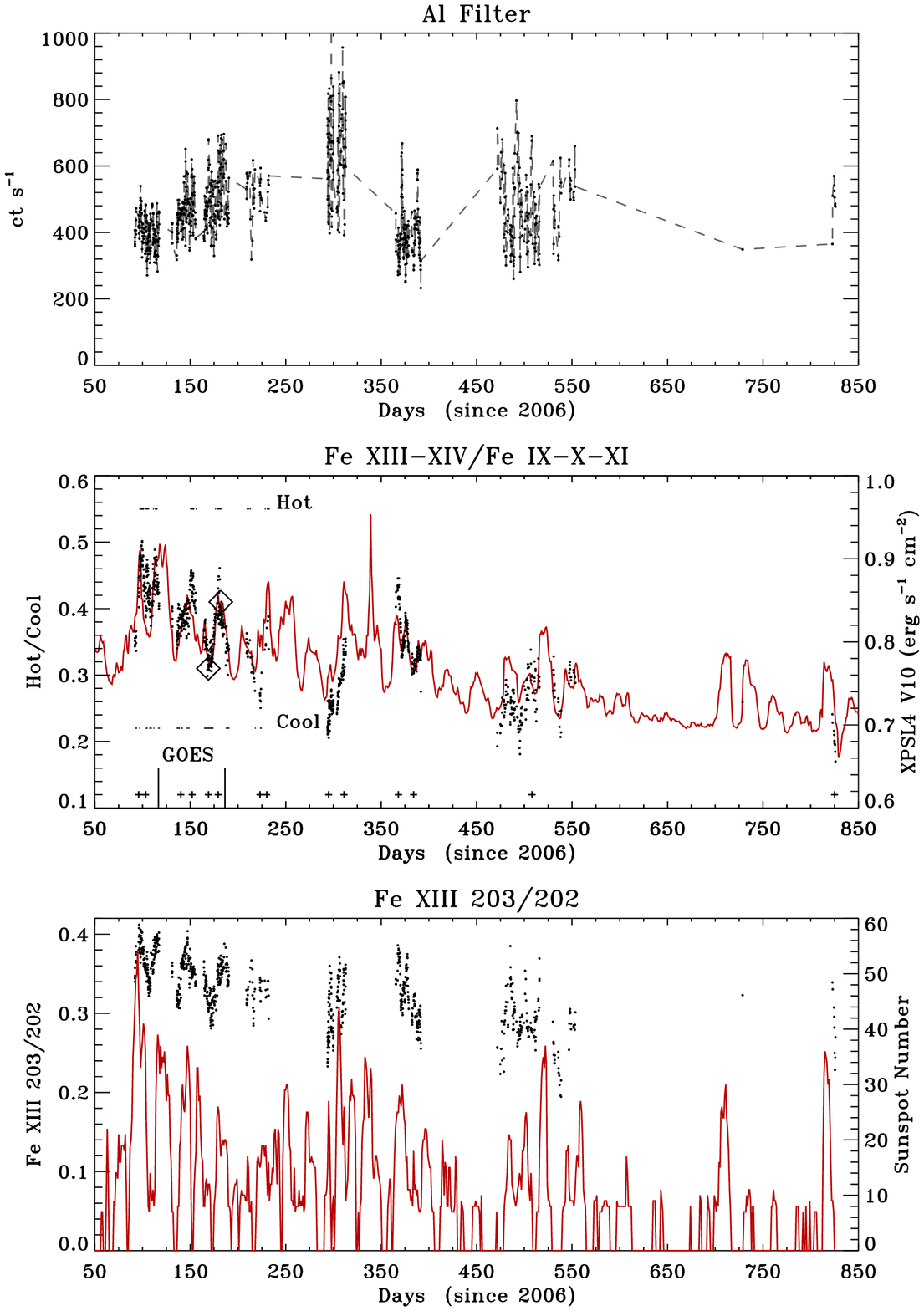}}
 \caption{Top: Integrated Al-filter lightcurve (170\rng273\AA).
  Middle: Hot-to-cool ratio
 (dots) compared to SORCE XPSL4 model flux integrated over the CHIPS passband
(red line). The diamonds
 denote the two extreme times chosen for the SOHO/EIT
images depicted in Figure~\ref{F-soho}. Horizontal dashes denote the times
from which the hot (upper dashes) and cool (lower dashes) spectra were
created.  The times of the two GOES flares analyzed in Section~\ref{S-flare}
 are marked with vertical dashes.
The times corresponding to the extreme-case hot and cool spectra atlas
(Figure~\ref{F-specatlas}) are demarcated with + symbols.
Bottom: CHIPS density-sensitive
\fetwoothree /\fetwootwo\ line ratio (dots) and the International
 Sunspot Number (red line).
 For all three panels, only observations showing less than 12\% multi-channel
 contamination are shown.}
 \label{F-lightcurves}
 \end{figure}

The integrated Al-filter count rate is determined over the passband
170\rng273 \AA\ and is plotted for the two year period 
as Figure~\ref{F-lightcurves} (upper panel).
The median count rate is 447 ct s$^{-1}, +155, -135$.
The large scatter from one observation to the next is greater than
the variations seen in the integrated XPSL4 lightcurve
(Figure~\ref{F-lightcurves}, red line, middle panel) and
is probably caused by the satellite pointing error of $\pm 1$\degree\
which affects the throughput of the light-leak.
% The Zr filter lightcurve tracks the Al filter very closely.
% However, inspection of detector images show non-uniform illumination
% along the slit with the Zr filter ocasionally being as bright
% as the Al, and free of contamination. 
Thus, any given CHIPS solar observation is only good to within a factor
of $\approx$ two in absolute flux.
Line ratios and correlations, however, are much more reliable since
systematic uncertainties and instrumental effects cancel out.

%% Figure lightcurves.eps
 
The variations in count rate observed in Figure~\ref{F-lightcurves}
(upper panel)
may be caused by a scale factor that is constant over the passband,
or may also have a wavelength dependence.
If the scatter is caused by just a scale factor, then it will cancel out
when taking line ratios.
If, however, there is also a time-varying wavelength
dependence in sensitivity,
then the degree of correlation in line intensities should decrease as the
separation in wavelength between a pair of lines increases.

\begin{figure} 
 \centerline{\includegraphics[width=1.1\textwidth,clip=]{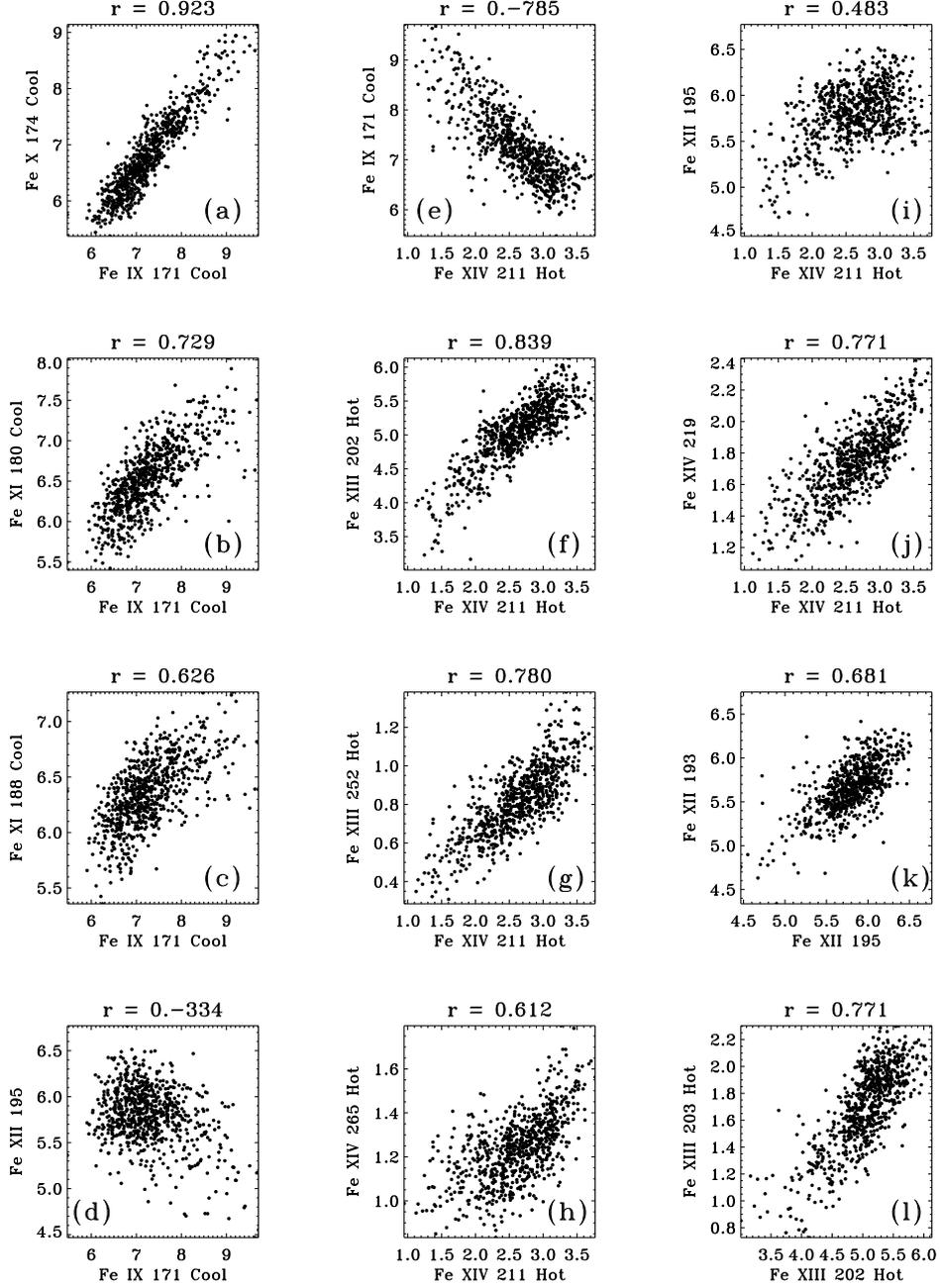}}
 \caption{Correlations of Fe-line intensities in units of percent of total
  Al-filter countrate. Cool {\it vs.} Cool (a\rng c). Cool {\it vs.} Hot (e).
 Hot {\it vs.} Hot (f\rng h,j).
 The Fe {\sc xii} 193 and Fe {\sc xii} 195 lines are poorly correlated
  with the Cool (d), and Hot (i) groups, and as well as with each other (k). 
  The density-sensitive Fe {\sc xiii} 203 and Fe {\sc xiii} 202 lines are
  moderately correlated (l).}
 \label{F-correlations}
 \end{figure}

To determine line intensities, the scale factor needed to match the spectrum in
question to the reference spectrum is first determined.
This scale factor is then applied to the mean background reference spectrum
and each spectrum is then background subtracted.
Count rates are next determined in 2 \AA\-wide bins centered on the
wavelengths of the desired features.
For the correlation analysis, the individual line intensities 
for each observation are presented as a percentage of
total count rate of the entire passband for that observation. 
Correlation plots are made that compare lines of both the same and different
ionization states, and lines both close and far apart in wavelength.
Some of the more instructive cases are presented as 
Figure~\ref{F-correlations}, and fall into three distinct categories.
The first group are the strongly correlated lines of Fe {\sc ix}
through Fe {\sc xi} (dominated by \feone)
which peak in brightness at a temperature around 1 MK
(Figure~\ref{F-correlations}(a\rng c)).
The second group are the 2 MK ionization states of Fe {\sc xiii} and {\sc xiv}
(dominated by \feeleven) which
also show strong correlation in Figure~\ref{F-correlations}(f\rng h, j).
Note that the  first two groups are distinctly anti-correlated
(Figure~\ref{F-correlations}(e)).
When the 1 MK lines are bright, the 2 MK lines are faint, and {\it vice versa.}
The third set involves \fethree\ and \fefive, which correlate poorly
with either of the first two groups (Figure~\ref{F-correlations}(d,i)).
Lastly, there is a set of lines (not shown in Figure~\ref{F-correlations})
that show bimodal distributions
({\it i.e.}, both positive and negative correlations simultaneously).
We interpret these cases as involving blends of both high and low temperature
lines ({\it e.g.}, Fe {\sc xiv} and Si {\sc ix} at 227 \AA, Figure~\ref{F-alspec}).

%% Figure corr.eps
 
If the variations in intensity between observations are caused by a change in
the wavelength dependence of the throughput, then {\it all} lines in a
given wavelength region would vary in unison regardless of ionization state.
This is clearly not the case for these data.
The observed variations in count rate of the CHIPS spectra when
features from one ionization state are compared to another
are caused by real temporal variations in plasma temperature.
This effect is best illustrated by combining the count rates of a few of the
strongest lines into two temperature groups:
The 1 MK ``cool'' group consists of \feone, \fefour, \feeighty, and \feeight,
while the 2 MK ``hot'' group of \fetwootwo, \feeleven, \fefiftytwo, and
\fesixtyfive.
The ratio of the hot-to-cool group count rates is plotted as
Figure~\ref{F-lightcurves} (dots, middle panel).
Any uncertainty in the exposure time or throughput cancels out
when taking this ratio.
For comparison, the XPSL4 flux is also plotted (red line, middle panel).
Evident in the hot-to-cool ratio is the 27-day mean rotation period of the Sun
(manifested as 27.2-day power in a Fourier spectrum of the hot-to-cool
ratio lightcurve),
the general decrease in high-temperature activity, and
the correlation in slope of the two satellite lightcurves
({\it i.e.}, when CHIPS sees an increase in temperature,
the XPSL4 spectral model shows an increase in flux).
However, when the same line ratios are extracted from the XPSL4 spectral model
and compared to the CHIPS ratios, the correlation vanishes.
This fact indicates that CHIPS is seeing a real change in plasma temperature
caused by solar rotation, while the XPSL4 model is just reflecting a
change in overall intensity.
In the bottom panel of Figure~\ref{F-lightcurves} we plot the
density-sensitive line ratio \fetwoothree / \fetwootwo\ (which is
discussed in Section~\ref{S-modeling}),
and the International Sunspot Number (red line).
Note that the slopes of the CHIPS hot-to-cool ratio, XPSL4 model, sunspot
number, and Fe 203/202 ratio are all positively correlated.

%% Figure hot_minus_cool.eps
 
 \begin{figure} 
 \centerline{\includegraphics[width=1.1\textwidth,clip=]{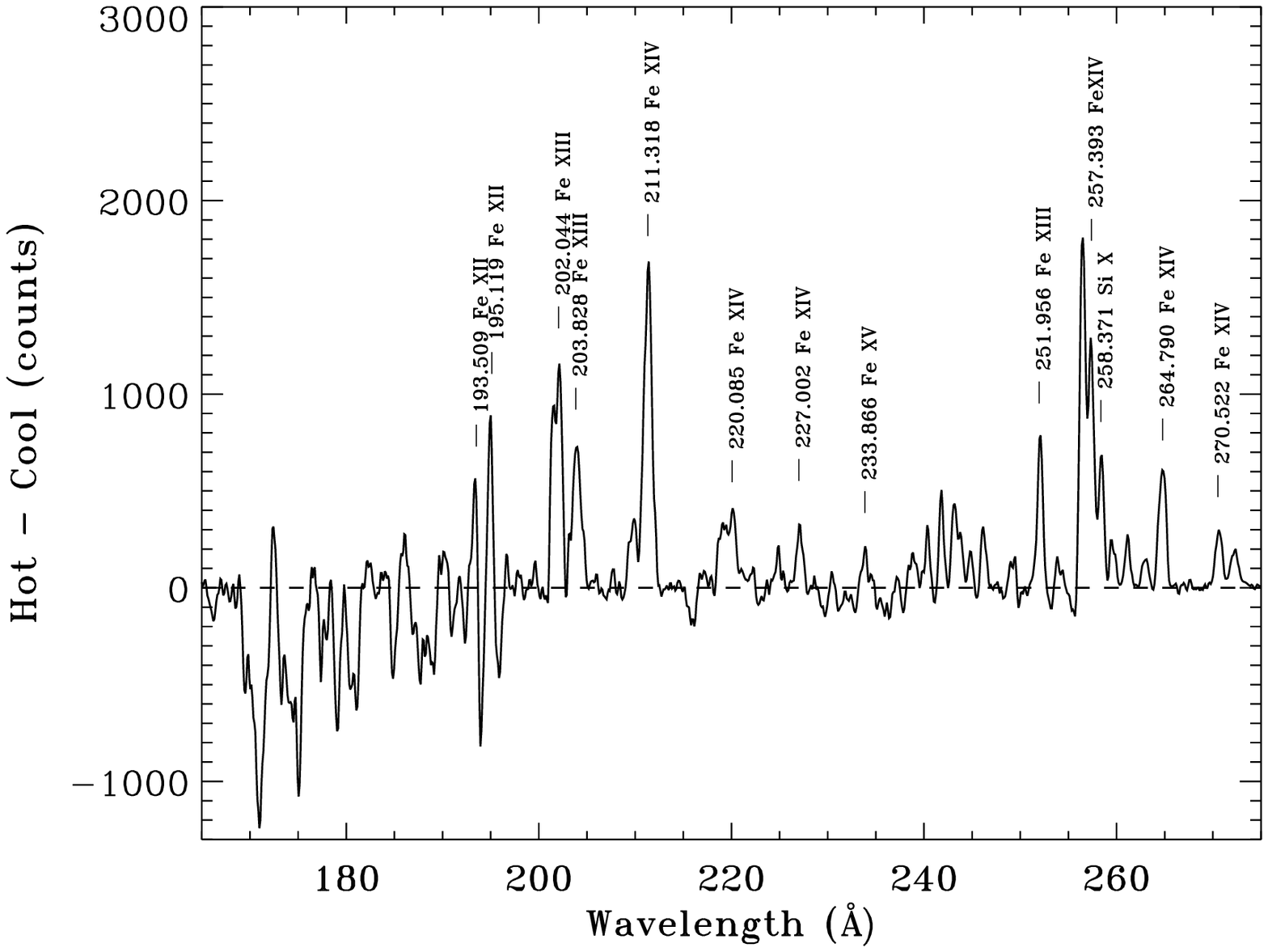}}
 \caption{Residuals of the ``hot'' 2 MK minus ``cool'' 1 MK spectra.
 The times from which the hot and cool spectra are created are marked
with horizontal dashes in Figure~\ref{F-lightcurves} (middle). 
Only the higher ionization states Fe {\sc xii} to Fe {\sc xv} show a significant positive residual.  The large negative residuals shortward of 190 \AA\ are
from the six bright 1 MK Fe {\sc ix\rng xi} lines.}
 \label{F-hotminuscool}
 \end{figure}

To further show that the temporal variations in temperature are real,
we construct two spectra filtered on times when the hot-to-cool ratio is
at maximum or minimum during 2006 (a period when the XPSL4 lightcurve
is relatively flat, Figure~\ref{F-lightcurves} (middle)).
The ``hot'' spectrum minus the ``cool'' spectrum is presented as
Figure~\ref{F-hotminuscool} and shows that only lines with
a high ionization state (Fe {\sc xii} to Fe {\sc xv}) show a positive residual.
Thus, we conclude that the observed hot-to-cool ratio temporal variations
are dominated by differences in plasma temperature and not 
some instrumental effect, and that,
because of the observed 27.2-day period, are caused by variations in emission
from hotter and cooler regions rotating into, and out of, view.
This behavior has also been seen by the TIMED/SEE instrument \cite{Woods05}.

\begin{figure} 
 \centerline{\includegraphics[width=0.75\textwidth,clip=]{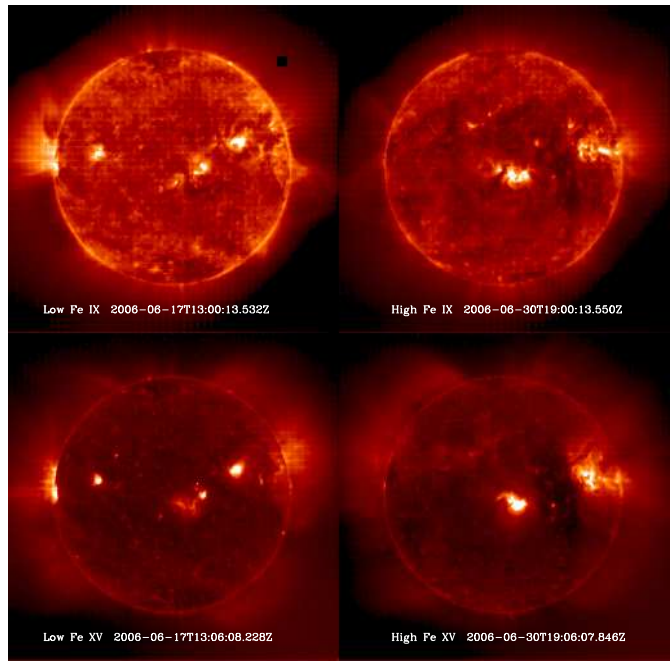}}
 \caption{SOHO/EIT images for 171 \AA\ and 284 \AA\ channels for the
 minimum (left), and maximum (right) hot-to-cool ratio indicated by diamonds 
in Figure~\ref{F-lightcurves}.}
 \label{F-soho}
 \end{figure}

We also chose two times about 13 days apart
($\approx 1/2$ the solar rotation period) where the hot-to-cool
ratio was at minimum or maximum in 
Figure~\ref{F-lightcurves} (diamonds, middle panel),
retrieved the corresponding SOHO/EIT full-disk images for the
171 \AA\ and 284 \AA\ channels,
and present them as Figure~\ref{F-soho}.
Visual inspection of the images shows no obvious changes in the number
and intensity of active regions.
However, several independent data sources show an increase in solar activity
during this period. The ratio of the SOHO images (284 \AA\ / 171 \AA)
increased by 25\%, the hot-to-cool ratio of CHIPS increased by 50\%,
and the CHIPS Fe {\sc xiii} 203/202 ratio by 39\%.
The sunspot number increased from zero to around 20 during this 13-day period,
and the ratio of {\it Michaelson Doppler Interferometrer} \cite{Scherrer95}
images extracted for these two times\break 
{\tt (mdi\_fd\_2006.06.17\_12:51:30.fits, mdi\_fd\_2006.06.30\_17:36:30.fits)}\break
shows an increase in magnetic flux by a factor of 1.65
which implies an increase of EUV and X-ray flux \cite{Pevtsov03,Fludra08}.
The XPSL4 model shows an increase in total flux of 8\% in the CHIPS passband.
These changes are all in the same direction
and indicate an increase in solar activity over the 13-day period.
The positive correlation between the CHIPS and SOHO/EIT line ratios
suggest the possibility of using the CHIPS spectra to help calibrate
the SOHO/EIT 171 \AA\ and 195 \AA\ broadband images,
which in turn could be used as proxies to
better calibrate other instruments for time periods not covered
by CHIPS.  We explore these ideas further in Section~\ref{S-Conclusions}.

%% Figure soho_color.eps
 
    \subsection{Daily Averaged Spectra}
      \label{S-dayspec}

\begin{figure} 
 \centerline{\includegraphics[width=1.0\textwidth,clip=]{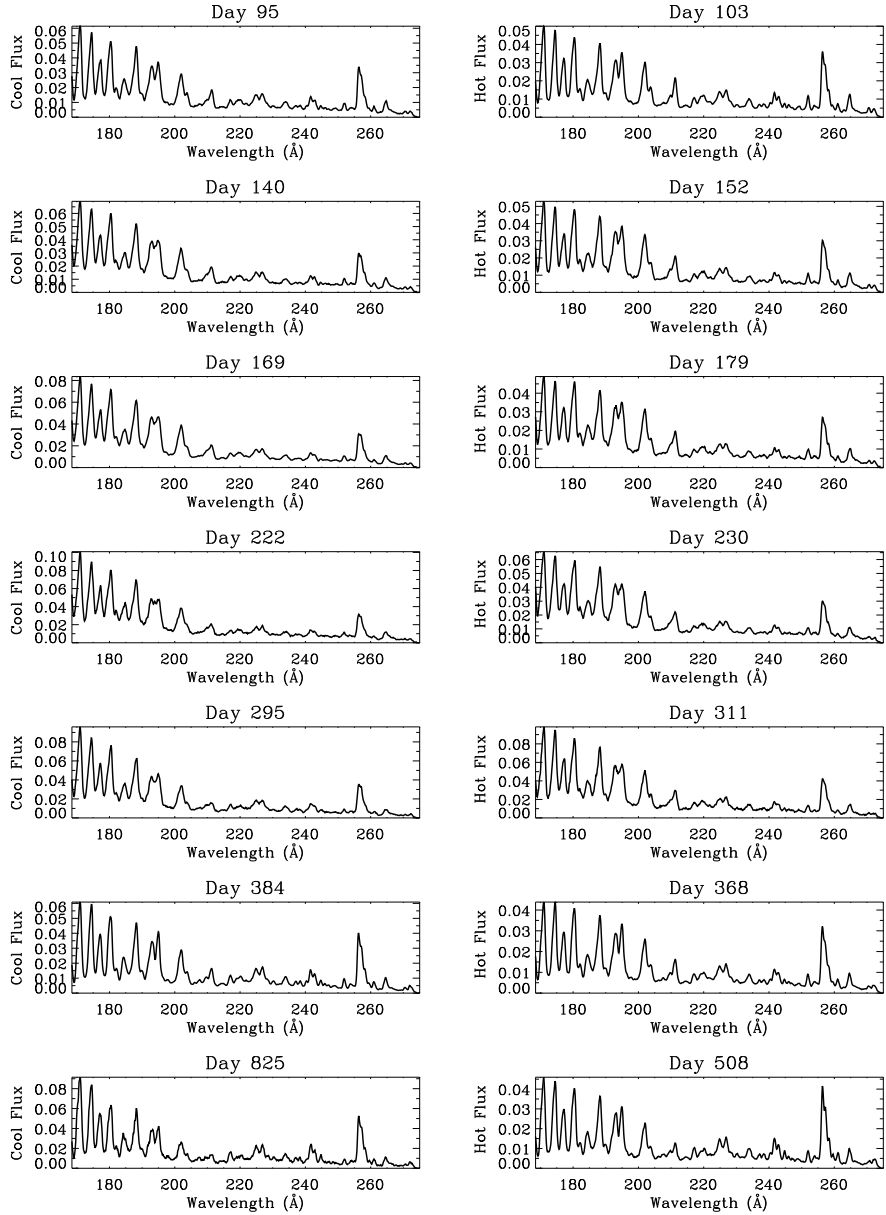}}
 \caption{Daily averaged Al-filter spectra covering two years for times 
 corresponding to local extrema of the hot-to-cool ratio 
 (marked with +symbols in Figure~\ref{F-lightcurves}, (middle)).
 Cooler spectra are on the left, hotter on the right in
 units of (erg s$^{-1}$ cm$^{-2}$ \AA$^{-1}$) which are accurate
 to within a factor of two.}
 \label{F-specatlas}
 \end{figure}

The significant variations seen in Figure~\ref{F-lightcurves} are the
short-period ($\approx$ 14 days) changes in the hot-to-cool line group ratio
and the general decrease in intensity of the 2 MK lines over the two year span
of the observations.
For reference, we present an atlas of extreme-case hot and cool spectra
%as
in Figure~\ref{F-specatlas} that illustrates the transition from moderately
active Sun to quiet Sun (compare day 103 to day 825).
The hot and cool spectra were extracted during times corresponding to
local maxima and minima of the hot-to-cool ratio indicated with
+ symbols in Figure~\ref{F-lightcurves} (middle panel). 
These spectra are not background-subtracted and so illustrate
the typical quality of the CHIPS multi-channel contamination-free
daily averaged spectra.    

    \subsection{CHIANTI Modeling}
      \label{S-modeling}

The CHIPS EUV spectra are from the full solar disk, and as such are the
combination of plasma features from the corona and transition region
which exhibit large ranges of temperature and density.
The dominance of the Fe emission lines, however, allows us to decompose
the spectra into several principal components.
The overriding strategy is to adequately fit the spectra with the fewest
number of atomic elements, temperatures, and densities, and
proceeded in this order: identify lines, the peak temperatures
associated with the observed lines, and the densities dictated by
specific line ratios.

Starting with just Fe, the bulk of the EUV emission is accounted for by
a 1 to 2 MK plasma.  In this temperature range, a few more features are
accounted for by Si and Ni.
The residual flux indicates the presence of O and He, but at lower
temperatures of \logt = 5.0 to 5.5.
The best density diagnostic available at the CHIPS spectral resolution is the
line ratio \fetwoothree/\fetwootwo\ whose average value is $0.34 \pm 0.05$
for the entire mission and varied in step with the hot-to-cool ratio
({\it i.e.}, higher densities are observed when the spectra are hotter,
dots on Figure~\ref{F-lightcurves}, middle and lower panels).
This ratio gives an electron density range of \logten$N_e$ (cm$^{-3}$) =
8.5 to 8.7 which is at the low end of the typical densities determined
in and around active regions by the {\it Hinode}/EIS
\cite{Young09,Warren08}.
%%is one to two orders of magnitude lower than 
Fainter lines from other elements
(such as Ne, Mg, S, Ar, and Ca
seen by the {\it Solar EUV Rocket Telescope and Spectrograph} (SERTS):
 \opencite{Thomas94})
%from a slit spectrogram of an active region 
were searched for, but could not be identified with any certainty.
The spectral resolution of CHIPS is the limiting factor,
not the instrument sensitivity.

The standard Differential Emission Measure (DEM) models 
(quiet sun, active region, coronal hole) that come with the
CHIANTI plasma package version 5.2.1 \cite{Landi06,Dere97} did not
fit the CHIPS spectra very well, either singly, or in combination.
We instead built a set of isothermal plasmas combining the elements
Fe, Ni, and Si (using the solar coronal abundances of \inlinecite{Feldman92}) 
at temperatures ranging from \logten$T$ = 
5.8 to 6.4 at 0.1 intervals.
We chose an average density of \logten$N_e$ (cm$^{-3}$) = 8.6
since the values determined from observations that ranged from 8.5 to 8.7 
made insignificant differences in the model spectra.
An isothermal plasma of He at \logten$T$ = 5.0,
and three of O at \logten$T$ = 5.3, 5.4, and 5.5 were also constructed,
all at a density \logten$N_e$ (cm$^{-3}$) = 10.0 typical
of the transition region \cite{Doschek91}.
Our model is simply the linear combination of these individual plasmas
with the only free parameters being the emission measures.

Because the changes in the CHIPS spectra are large over the short time 
period of one-half solar rotation, but minor over longer periods,
we built two time-averaged spectra corresponding to
the minimum and maximum time intervals of the hot-to-cool ratio of
Figure~\ref{F-lightcurves} (middle panel) during 2006.
We fit these hot and cool spectra separately.
Good fits were found requiring plasmas of \logten$T$ = 6.0, 6.1, and 6.3.
Inclusion of the He plasma decreased $\chi^2$ significantly, while that of
O decreased $\chi^2$ by 5 to 8\%.
The inclusion of the \logten$T$ = 5.8, 5.9, 6.2, and 6.4 temperature components
of the Fe-Si-Ni plasma did not reduce $\chi^{2}$ significantly, and thus
are not included in the final CHIANTI model fits.

%% Figure chianti_fits.eps

 \begin{figure} 
 \centerline{\includegraphics[width=1.1\textwidth,clip=]
 {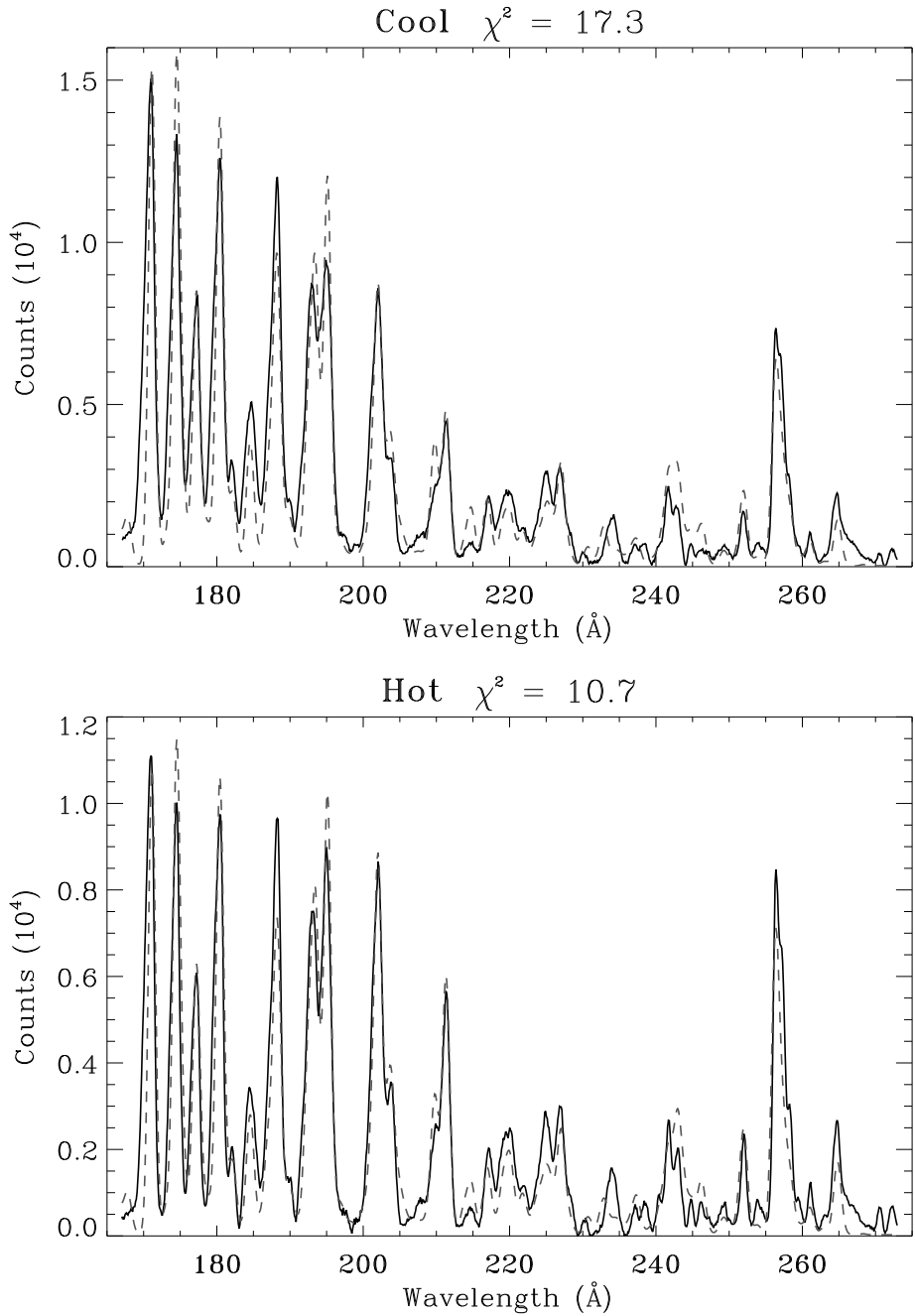}}
 \caption{CHIANTI model fits (dashed lines) to the Al-filter 1 MK ``cool''
 spectrum (solid line, upper panel)  and Al-filter 2 MK ``hot'' spectrum
 (solid line, lower panel). The reduced $\chi^{2}$ for each fit is also
 indicated.}
 \label{F-hotcool}
 \end{figure}

    \subsection{Light-Leak Efficiency}
      \label{S-lightleak}
The modeling outlined in Section~\ref{S-modeling} successfully reproduced
the majority of observed features in the CHIPS spectra, but showed
a systematic error in amplitude from one end of the spectrum to the other.
This residual is an expected consequence of the wavelength dependence
of the light-leak.
The exact nature of the light-leak cannot be modeled directly since the
angle of incidence and surface roughness of the Au and Cu/Be slit-shutter
mechanisms are poorly known.
However, the functional form is estimated using the
Lawrence Berkeley National Laboratory X-ray Properties
of Matter web utility \cite{Henke93}.
We find a smoothly varying function that increases in reflective efficiency
with wavelength, and which is well approximated by a simple exponential
over the wavelength range of interest
 \begin{equation}  \label{Eq-lightleak_eff}
     {\rm efficiency}(\lambda) = S \exp{(\alpha\lambda)}.
   \end{equation}
where $\alpha$ governs the wavelength dependence of the light-leak,
and $S$ is a scale factor, which, when combined with the CHIPS nominal
effective-area curves, converts count rate to physical units
(determined in Section~\ref{S-chips_vs_xps}).

Our final model is thus the sum of the individual plasmas multiplied
by an exponential.
The model fits to the cool and hot spectra are shown as Figure~\ref{F-hotcool},
and the corresponding emission measures presented in Table~\ref{T-table1}.
The random errors of the emission-measure determinations are $\approx 8\%$.
The total uncertainty is discussed in Section~\ref{S-Conclusions}. 
Since the He and O emissions are from the transition region
\cite{Reeves77} and are optically thick,  % find other reference?
the CHIANTI-derived values of emission measure may match the data,
but are probably only lower limits.

The CHIANTI model fits to the large Al and Zr filters produced essentially
identical values for temperature, density, and light-leak efficiency.
On average, the Zr data are of inferior quality suffering from
multiple-channel contamination and non-uniform slit illumination.
Furthermore, because the Zr passband (140\rng202 \AA) is smaller than
the Al filter, fewer lines are available to constrain the model parameters.
However, the presence of Ni lines in the Zr filter spectra clearly indicate
a plasma component of \logt = 6.3, and the relative weakness of the
Fe {\sc viii} features at 168 \AA\ indicate the absence of significant plasma
components of \logt = 5.8\rng5.9.
We present only the Al-filter results in Table~\ref{T-table1} as
they are the most reliable.
A new version of CHIANTI (6.0.1) was released in October 2009.
The changes, however, do not affect the main conclusions of this work.

\begin{table}
\caption{Emission Measures$^{a}$ (cm$^{-5}$) determined from \nobreak{CHIANTI}
 modeling of the Al-filter spectra for the different element and
 temperature plasma components.}
% for \logt = 6.0, 6.1, and 6.3.}
%l: left, c: center, r: right. 
%Using two \$s permits one to insert equation-like features (see last column). 
%The inclusion of $\sim$ adds a blank to approximately align the numbers
%of the last two columns (see the \LaTeX\ file).

\label{T-table1}
\begin{tabular}{lcccccc}     % define the column alignment
                           % l: left, c: center, r: right
\hline                     % horizontal line
 &  \multicolumn{3}{c}{Fe-Si-Ni} & He     & O     & \\
Spectrum & 6.0   & 6.1   & 6.3   & 5.0    & 5.5   & $\chi^{2}$\\
\hline

Cool     & 20.53 & 20.53 & 20.40 &  20.95 & 20.82 & 17.3 \\
Hot      & 20.36 & 20.45 & 20.54 &  21.05 & 20.63 & 10.7 \\

\hline
\end{tabular}

$^{a}$\logten\ of the emission measure and temperature are given.\\
 Relative uncertainties in the emission measures are typically 8\%.\\
 Absolute uncertainties are discussed in Section~\ref{S-Conclusions}.\\
 Ionization fractions used are from \inlinecite{Mazzotta98} and\\
 solar corona abundances from \inlinecite{Feldman92}.\\
\end{table}

    \subsection{Comparison to SORCE/XPS and SDO/EVE}
      \label{S-chips_vs_xps}
\begin{figure} 
 \centerline{\includegraphics[width=1.1\textwidth,clip=]
 {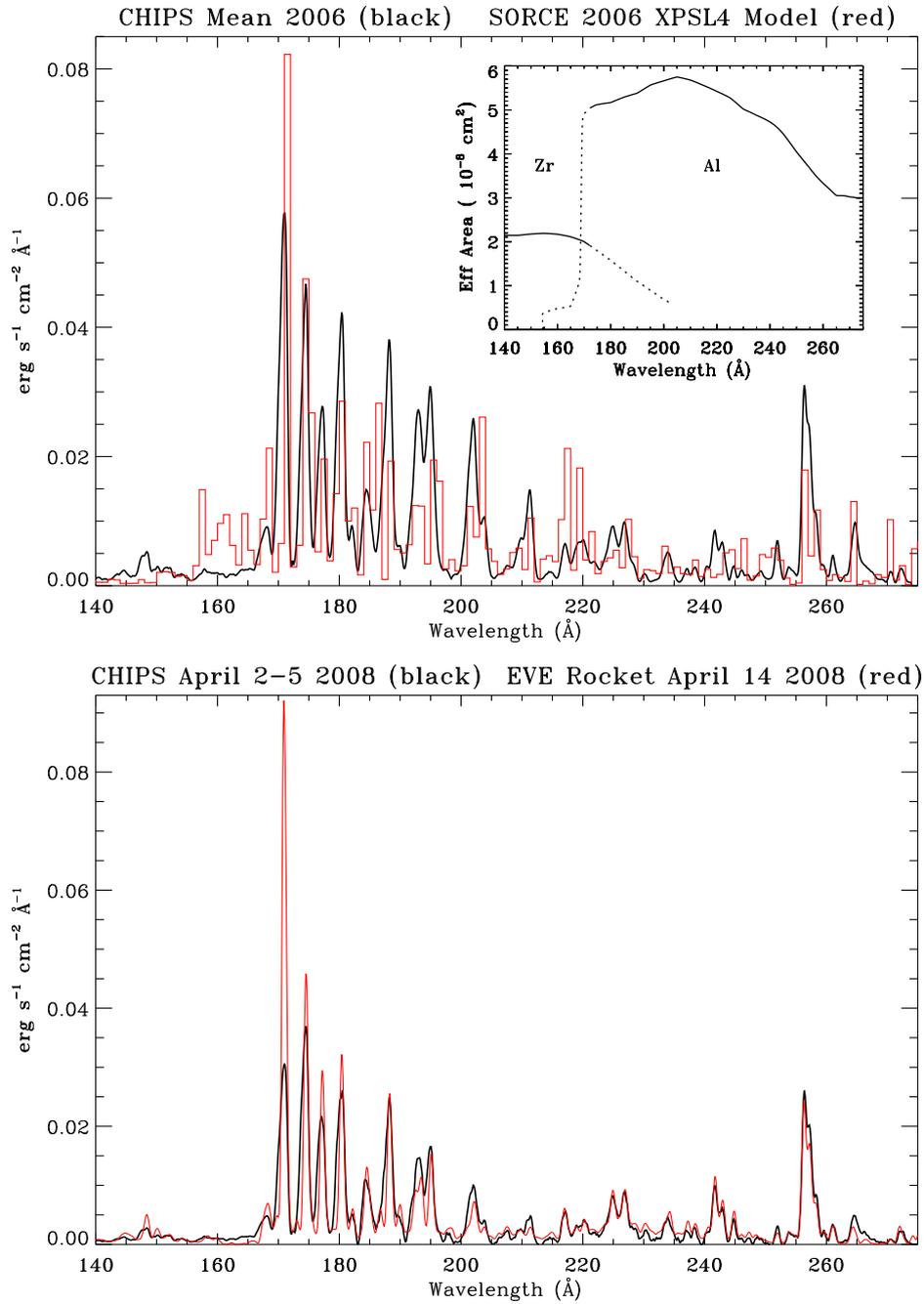}}
 \caption{Top: CHIPS spectrum binned at 0.10 \AA\ (black line) scaled to
 XPSL4 model binned at 1.0 \AA\ (red histogram).
 The inset plots the nominal effective-area curves of the CHIPS Al and Zr
 filters multiplied by the modeled efficiency curve of the light-leak.
 The solid segments of each curve are used for the flux determinations. 
 Bottom: CHIPS spectrum for
 2\rng5 April 2008 (black) scaled to EVE rocket spectrum of 14 April 2008 (red).}
 \label{F-chips-xps}
 \end{figure}

To answer specific questions regarding solar irradiance, solar variability,\break
differential-emission-measure maps, and other solar-physics problems, the\break
CHIPS spectra must be converted to physical units.
We accomplish this by determining the average scale factor $[S]$ in
Equation~(\ref{Eq-lightleak_eff}) by
comparing the CHIPS spectra to contemporaneous
SORCE/XPS Level 4 model fluxes \cite{Woods08etal} 
integrated over the passband 145-273 \AA.
We chose the first 370 days since 2006 because during this period
the XPSL4 flux remained fairly flat (see Figure~\ref{F-lightcurves}).
The CHIPS ground-based effective-area curves are multiplied by
Equation~(\ref{Eq-lightleak_eff}) ($eff(145)= 6.04 \times 10^{-7},\
eff(275) = 3.75 \times 10^{-6},\ eff_{{\rm mean}} = 1.73 \times 10^{-6}$).
We show the mean-flux calibrated CHIPS spectrum as 
Figure~\ref{F-chips-xps} (top)
and compare it directly to the XPSL4 spectrum extracted for the
same time period.
The spectral features are in general agreement, but there are several
significant differences which we discuss in Section~\ref{S-Conclusions}.

In Figure~\ref{F-chips-xps} (bottom) we show the EVE rocket spectrum
of 14 April 2008 \cite{Chamberlin09,Woods09}, and compare it directly to the 
scaled CHIPS spectra averaged over 2\rng5 April 2008.
The excellent agreement between the two spectra shows that both instruments are
free from spurious features such as grating ghosts and second-order effects,
and that the CHIPS
background subtraction and light-leak efficiency determination are reasonable.
The only significant difference is the intensity of the \feone\ line,
which we address in Section~\ref{S-Conclusions}.

%% Figure chips_compare.eps

    \subsection{Solar Flares}
      \label{S-flare}
\begin{figure} 
 \centerline{\includegraphics[width=1.1\textwidth,clip=]
 {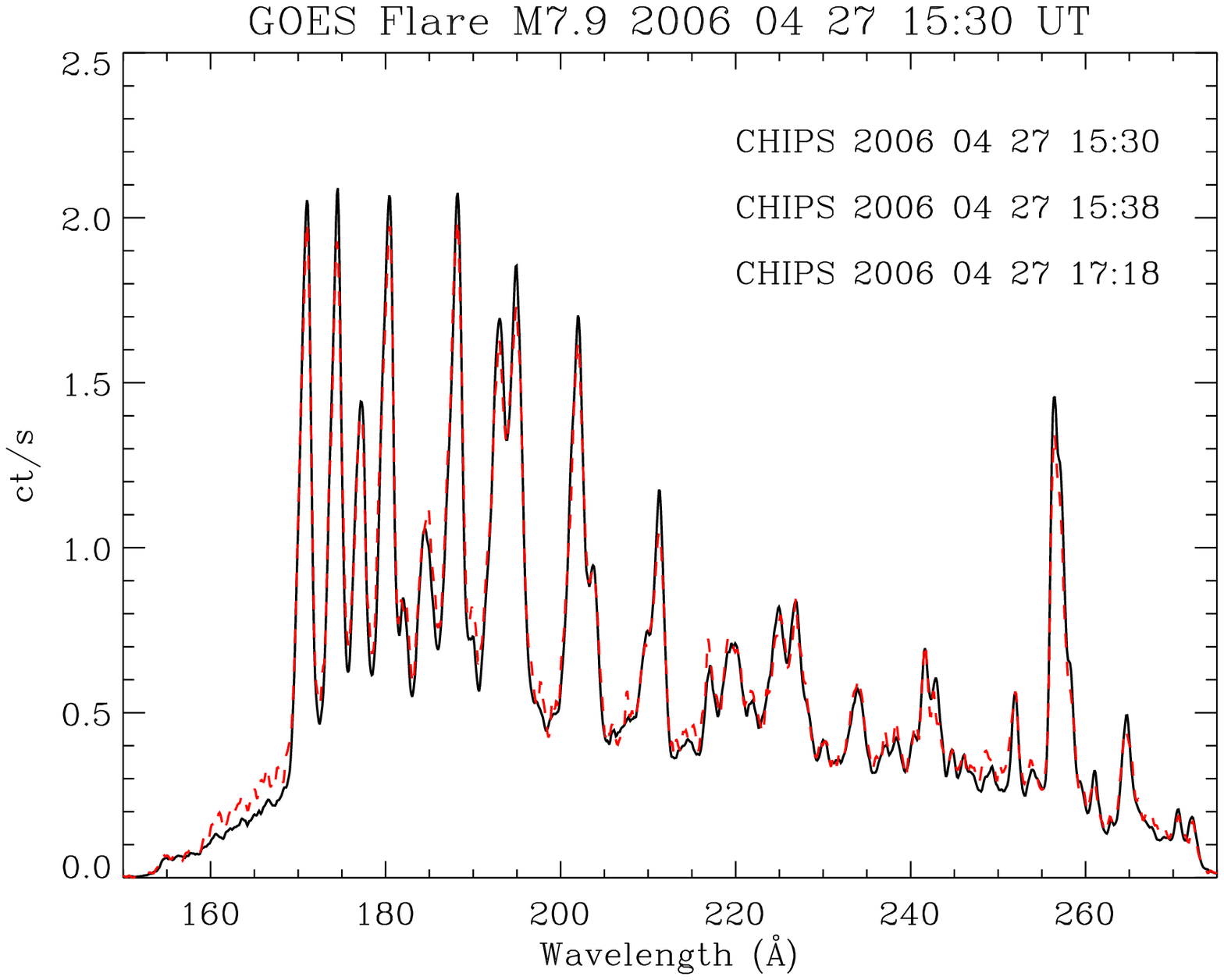}}
 \caption{Raw CHIPS spectrum summed over five days prior to flare binned at
 0.11 \AA\ (black) compared to sum of three observations during flare (red).
 The only significant difference is some minor multi-channel contamination
 visible in the 160 to 170 \AA\ region of the flare spectrum.}
 \label{F-flare}
 \end{figure}

During 2006, CHIPS observed an M-class flare,
and narrowly missed a second flare by 24 hours (GOES X-ray peak times
27 April 2006 15:22UT M 7.9, 06 July 2006 08:13UT M 2.5, respectively). 
A baseline spectrum was extracted over a five-day
period just prior to the first flare and compared to the three CHIPS
observations obtained during flare.
We present these spectra in Figure~\ref{F-flare} in their raw state which is
free from any possible systematic error introduced by background subtraction.
There is no evidence of a change in plasma temperature.
The CHIPS observations of the
second flare which were obtained 24 hours after the GOES X-ray peak also 
showed no differences between the pre- and post-flare spectra.

\section{Discussion and Conclusions}
     \label{S-Conclusions}

CHIPS recorded spectra of the solar disk between 140 \AA\ and 273 \AA\
over a two-year period.
The general appearance of the spectrum remained relatively stable.
The emission lines labeled in Figures~\ref{F-alspec} and ~\ref{F-zrspec}
did not appear and disappear, but were visible at all times, even during 
moderate (M-class) flares and the quiet-Sun period of April 2008.
There was, however, variation in the 
ratio of the hot and cool emission line groups of about $\pm$ 25\%,
and a variation in the 1 to 2 MK plasma density of $\pm$ 15\%.

The CHIPS spectra indicate a 1 to 2 MK plasma dominated by Fe emission.
Iron is also the most significant element for higher temperature spectra of
active regions and flares,
and for wavelengths both shorter and longer than the CHIPS passband
(see the rocket-flight spectra of \opencite{Behring72};
\opencite{Malinovsky73}; \opencite{Thomas94};
and the {\it SkyLab} data of \opencite{Dere78}).
In Table 2 we present the fraction of total intensity for each element used
in the CHIANTI fits, and the temperature of peak emission.
Recent theoretical work by \inlinecite{Colgan08} using the Los Alamos ATOMIC
plasma code concludes that Fe plays the dominant role in total radiative loss
of the corona, even when it is depleted relative to the other elements. 

\begin{table}
\caption{Fraction of total intensity by species, and temperature
 of peak emission between 140 \AA\ and 273 \AA.}
\label{T-table2}
\begin{tabular}{ccl}       % define the column alignment
                           % l: left, c: center, r: right
  \hline                   % horizontal line
Element & Fraction (\%)   &  \logten\ T$_{{\rm peak}}$\\
  \hline
Fe      & 85.3 & 6.0\rng6.3 \\
Si      & ~5.7 & 6.1 \\
Ni      & ~1.8 & 6.3 \\
He      & ~4.3 & 5.0 \\
~O      & ~3.0 & 5.5 \\
  \hline
\end{tabular}
\end{table}

The CHIPS hot and cool spectra are reasonably well modeled with a small
set of isothermal plasmas of just 
five elements, five temperatures, and two densities.
The standard DEMs supplied with the CHIANTI package do not model the
CHIPS spectra nearly as well.
These results are in agreement with those of \inlinecite{Cirtain06} who,
in a study of the unresolved EUV corona, conclude that above non-flaring active
regions the corona is isothermal (typically between 1 and 2 MK),
has an average electron density of \logten$N_e$ (cm$^{-3}$) = 8.0,
and varies little in time.
In a subsequent survey of 20 active regions,
J.~Cirtain (2010, private communication)
also finds that the unresolved EUV corona component comprises 55\% of the
total EUV emission.
We propose that the variation seen by CHIPS in plasma temperature
over the solar rotation period is caused by hotter unresolved plasma 
associated with active regions, and cooler plasma rotating into
and out of view.

The spectral shape, line ratios, temperatures, densities, and light-leak
efficiency presented here are determined directly from the CHIPS data.
The integrated XPSL4 flux is used only to adjust the CHIPS spectra by a single
scale factor (one per observation) when converting count rate to physical
flux units.
Any CHIPS solar spectrum, when divided by the exposure time and the
modified effective-area curves, now yield a calibrated flux
(erg s$^{-1}$ cm$^{-2}$ \AA$^{-1}$)
to within a factor of two if raw, or to within 50\% if scaled to XPSL4
\cite{Woods08etal,Woods10}.
However, since the XPS CHIANTI model spectrum shows several significant
differences when compared to the empirical CHIPS mean solar spectrum,
there are probably additional systematic uncertainties in the integrated
XPSL4 flux.
In particular, the XPSL4 model under-predicts He {\sc ii}, and over-predicts
Ni {\sc xii} and Ni {\sc xiii} around 160 \AA.
Several other spectral features in Figure~\ref{F-chips-xps} differ
by factors of $\approx$ two.
The Ni {\sc x} and  {\sc xi} lines (seen by CHIPS shortward of 150 \AA)
are not yet in the CHIANTI (version 5.2.1) database.
In the XPSL4 spectra, \feone\ is typically twice as bright as its
closest neighbors,
whereas the CHIPS spectra consistently show it to be only $\approx$ 8\% brighter.
The XPSL4 line ratio \fetwoothree/\fetwootwo\ averaged 2.0 in 2006 which
is about a factor of six greater than the value determined from the
CHIANTI fits to CHIPS spectra for the same time period.
This large discrepancy implies that the electron density used by the
XPSL4 model is an order of magnitude larger than that required
to model the CHIPS spectra (see Figure 20 in \inlinecite{Young09}). 
In general, features visible at temperatures $>$ \logt = 6.3, and
$<$ \logt = 6.0, are over-estimated by the XPSL4 model.
One important difference to bear in mind between the CHIPS empirical spectra
and SORCE XPS Level 4 model is that XPSL4 is based on a single photodiode
(Ti filter, 1\rng70 \AA\ passband, \inlinecite{Woods08etal}),
which does not overlap the CHIPS spectral range.
So, there is no surprise that there are some differences.
These differences may result in systematic uncertainties in the XPSL4 model
flux and can be eliminated if the SORCE XPS
diode currents are used to scale
a CHIANTI model that is based upon DEMs derived from empirical spectra 
rather than assumed DEMs and out-of-band proxies.

%% Ratio of EVE / XPSL4 Version 10 = .679

The agreement between CHIPS and EVE  seen in Figure~\ref{F-chips-xps}
(bottom) is excellent apart from the factor of two discrepancy at 171 \AA.
A likely cause for this difference may be that the resolution of the 
effective-area curve of one of the instruments does not match the actual
instrument resolution in the region of the Al absorption edge around 169 \AA.
To test this idea we compare the \feone\ line to the \fefour\ line
since the two lines are close together in wavelength and, hence,
the difference in instrument throughput is small.
Over the course of the two year CHIPS mission, the 171/174 \AA\ line 
ratio averaged 1.09 $\pm$ 0.05, and 1.07 $\pm$ 0.09 for
the Al and Zr filters, respectively.
Because the Zr filter has no absorption edge near 171 \AA\
(see Figure~\ref{F-chips-xps}, inset),
we believe the ratio of near-unity is quite robust.
A quiet-Sun spectrum obtained in early 1997 from SOHO/GIS shows a 
171/174 \AA\ line ratio of 1.3 (see Figure 2 of \opencite{Kuin07}),
and the full-disk rocket spectrum of 04 April 1969 
(obtained during a period of some solar activity) 
shows a ratio 1.0 \cite{Malinovsky73}.

The total uncertainty in absolute flux of the CHIPS spectra is the sum of the
XPS absolute photometric calibration uncertainty (50\%)
and the systematic error introduced in scaling
the CHIPS spectra to the XPSL4 model. 
By definition, this latter error is zero when averaged over the entire
CHIPS band and over an entire solar-rotation period,
but as can be seen in Figure~\ref{F-chips-xps} (top) is often
a factor of $\approx$ two for individual features.
Since the XPSL4 products are at 1.0 \AA\ bin width, 
a quantitative line-by-line comparison
between CHIPS and XPSL4 is currently not possible.

The CHIPS spectra can be used to help cross-calibrate TIMED/SEE,\break
SORCE/XPS, and SOHO/EIT for periods of simultaneous observations.
Then SOHO/EIT line ratios could be used to determine relative amounts
of hot (2 MK) and cool (1 MK)
plasma to make a more realistic solar spectrum for the SORCE/XPS CHIANTI
models for periods where CHIPS data do not exist.

In summary, our primary findings are:

\begin{itemize}

\item Near simultaneous spectra from CHIPS and the SDO/EVE rocket show
     excellent agreement indicating that both instruments perform as designed,
      and that the CHIPS light-leak throughput is calibrated correctly.

\item Full-solar disk EUV spectra obtained April 2006 through April 2008
      show temporal variations of up to 25\% with a period of 27.2 days.
      CHIANTI model plasma fits to the spectra show changes in the relative
      proportions of 1 MK and 2 MK plasmas, and changes in plasma density
      of \logten$N_e$ (cm$^{-3}$) = 8.5 to 8.7 over this period.  

\item The CHIPS spectra are well modeled with a simple set of five isothermal
      plasmas requiring five temperatures, five elements, and two densities.
      A 1 to 2 MK plasma of Fe accounts for 85\% of the observed flux.

\item The full-disk spectra are less well fit by standard CHIANTI DEMs
      which over-predict features above \logt = 6.3 and
      below \logt = 6.0. This is true for individual spectra
      as well as those averaged over many solar rotations and suggests
      that the true solar DEMs are not smoothly varying functions.

\item The CHIPS spectra may be utilized to help cross-calibrate 
      contemporaneous SOHO/EIT, TIMED/SEE, and SORCE/XPS broad-band data,
      which ultimately will improve our knowledge of the total solar
      irradiance.

%%\item CHIPS observations of the Sun during occultation by the Earth limb 
%%      provide a promising avenue for probing the density of thermospheric 
%%      N${_2}$ and O${_2}$ at altitudes between about 170 and 280 km.

\end{itemize}

%instruments is that
%the SORCE/XPS diode response curve decreases over 2 orders of magnitude
%over the 170 to 260 \AA\ passband (Figure 2 in \opencite{Woods08etal}),
%while that of CHIPS is nearly flat (see Figure~\ref{F-chips-xps}, inset).
%Thus, the XPS diode count rates are dominateded by the
%shorter wavelength lines in this passband.  This is true for the diode
% that overlaps CHIPS, but the XPSL4 model does not use that diode!!!

%%%%%%%%%%%%%%%%%%%%%%%%%%%%%%%%%%%%%%%%%%%%%%%%%%%%%%%%%%%%%%%%%%%%%%%%%%
%% Appendix

\appendix   

All CHIPS solar observations will be made available through a public archive
a year from the acceptance of this paper.
Preliminary mean ``hot'' and ``cool'' state spectra may be found at
\url{http://ssl.berkeley.edu/chips/archive.html}.
Each observation is stored in standard FITS table format.
The polyamide-B/Al and Zr/Al-filter spectra are presented 
on a common, uniform (0.1 \AA) wavelength grid as both raw counts,
and in flux calibrated units.

The FITS headers contain spacecraft parameters determined
at the time of telemetry processing, plus subsequent parameters created
during the standard science pipeline processing.  Additional parameters unique
to the solar observations are also included.
Because the EUV emitting region of the Sun extends beyond the solar limb
and is quite variable, the Sun is treated as a point source.
Hence, flux units of (erg s$^{-1}$ cm$^{-2}$ \AA$^{-1}$) are used,
which are good to within a factor of two.

Finally, as a convenience to other researchers, the multiplicative
conversion factor required to scale the CHIPS spectra 
(over the 140\rng273 \AA\ passband) to the contemporaneous SORCE/XPS
Level 4 data products is also included for each observation.
However, if the XPSL4 products are recalibrated, then this scale factor
will also change somewhat.

%%%%%%%%%%%%%%%%%%%%%%%%%%%%%%%%%%%%%%%%%%%%%%%%%%%%%%%%%%%%%%%%%%%%%%%%%%%
%% Acknowledgements
%
 \begin{acks}
 This work is supported by the Office of Space Sciences, National Aeronautics
 and Space Administration, under Grant No. NAG5-5219.
 We thank Brian Welsch for numerous discussions and comments
 on the manuscript, the anonymous referee for many helpful improvements,
 Michael Sholl for Figures 1 and 2,
 Tom Woods and the LASP Team for providing the high resolution SDO/EVE
 rocket spectrum, and
 John McDonald, Jeremy Thorsness, and Mark Lewis of 
 the CHIPS Operations Team who kept the instrument
 flying for several years beyond its expected lifetime.
% Dr. Jerry Lumpe for assistance in analyzing the Earth limb occultation data,
\end{acks}

%%% %%%%%%%%%%%%%%%%%%%%%%%%%%%%%%%%%%%%%%%%%%%%%%%%%%%%%%%%%%%
%% Bibliography
%
% Using BibTeX
%
 \bibliographystyle{spr-mp-sola}
% %\bibliographystyle{spr-mp-sola-cnd} %% Alternative style: no title, no concluding page
 \bibliography{chips_solar}  
%
% Without BibTeX 
% \begin{thebibliography}{}
% \bibitem[\protect\citeauthoryear{Author}{Year}]{key}
%   <bibliographical entry>
%
% \bibitem[\protect\citeauthoryear{}{}]{}
%     
% \end{thebibliography}

\end{article} 
\end{document}